\documentclass[12pt, draftclsnofoot, onecolumn]{IEEEtran}

\ifCLASSOPTIONcompsoc
  \usepackage[nocompress]{cite}
\else
  \usepackage{cite}
\fi

\usepackage[utf8]{inputenc}
\usepackage{mathrsfs}
\usepackage{tabularx,lipsum,environ,amsmath,amssymb}
\usepackage[boxruled,linesnumbered]{algorithm2e}
\usepackage{cleveref}
\crefname{Lemma}{Lemma}{Lemmas}
\usepackage{booktabs}
\usepackage{algorithm2e}
\usepackage[table,xcdraw]{xcolor}
\usepackage{tabularx,lipsum,environ,amsmath,amssymb}

\usepackage{algorithm2e}
\usepackage{graphicx}            
\usepackage{amsmath}             
\usepackage{amssymb}
\usepackage{array}
\usepackage{layout}
\usepackage[margin=1.04in]{geometry}
\usepackage{amsmath,amssymb}

\makeatletter
\newsavebox\myboxA
\newsavebox\myboxB
\newlength\mylenA

\newcommand*\xoverline[2][0.75]{%
	\sbox{\myboxA}{$\m@th#2$}%
	\setbox\myboxB\null
	\ht\myboxB=\ht\myboxA%
	\dp\myboxB=\dp\myboxA%
	\wd\myboxB=#1\wd\myboxA
	\sbox\myboxB{$\m@th\overline{\copy\myboxB}$}
	\setlength\mylenA{\the\wd\myboxA}
	\addtolength\mylenA{-\the\wd\myboxB}%
	\ifdim\wd\myboxB<\wd\myboxA%
	\rlap{\hskip 0.5\mylenA\usebox\myboxB}{\usebox\myboxA}%
	\else
	\hskip -0.5\mylenA\rlap{\usebox\myboxA}{\hskip 0.5\mylenA\usebox\myboxB}%
	\fi}
\makeatother

\usepackage{etoolbox}

\makeatletter
\patchcmd{\@maketitle}{\raggedright}{\centering}{}{}
\patchcmd{\@maketitle}{\raggedright}{\centering}{}{}
\makeatother

\begin{document}

\title{Beam Entropy of 5G Cellular Millimetre-Wave Channels}

\author{\IEEEauthorblockN{Krishan K. Tiwari$^1$, \textit{SMIEEE}, Eckhard Grass$^{1,2}$, John S. Thompson$^4$, \textit{FIEEE}, and Rolf Kraemer$^{1,3}$ }
	\IEEEauthorblockA{$^1$IHP -- Leibniz-Institut f\"{u}r innovative Mikroelektronik, Im Technologiepark 25, 15236, Frankfurt (Oder), Germany.\\$^2$Humboldt-Universit\"{a}t zu Berlin, 10099, Berlin, Germany.~~~~~$^3$BTU-Cottbus, 03046, Cottbus, Germany.\\ $^4$IDCOM, School of Engineering, Edinburgh, EH9 3JL, U.K. \\
		Email addresses: {tiwari, grass, kraemer}@ihp-microelectronics.com, John.Thompson@ed.ac.uk}

}


\maketitle

\begin{abstract}

In this paper, we obtain and study typical beam entropy values for millimetre-wave (mm-wave) channel models using the NYUSIM simulator for frequencies up to 100 GHz for fifth generation (5G) and beyond 5G cellular communication systems. The beam entropy is used to quantify sparse MIMO channel randomness in beamspace. Lower relative beam entropy channels are suitable for memory-assisted statistically-ranked (MarS) and hybrid radio frequency (RF) beam training algorithms. High beam entropies can potentially be advantageous for low overhead secured radio communications by generating cryptographic keys based on channel randomness in beamspace, especially for sparse multiple-input multiple-output (MIMO) channels. Urban microcell (UMi) and urban macrocell (UMa) cellular scenarios have been investigated in this work for 28, 60, 73, and 100 GHz carrier frequencies and rural macrocell (RMa) scenario for 3.5 GHz.



\end{abstract}

\IEEEpeerreviewmaketitle

\section{Introduction}
 Global data traffic projections indicate that by 2022 the annual global internet protocol (IP) traffic will reach 4.8 zettabytes (ZBs) per year, traffic from wireless and mobile devices will account for 71 percent of total IP traffic, and mobile data traffic will increase sevenfold between 2017 and 2022 growing at a compound annual growth rate (CAGR) of 46 percent between 2017 and 2022, reaching 77.5 exabytes (EBs) per month by 2022 \cite{cisco}. 

Large scale MIMO systems in mm-wave and THz bands will play an instrumental role in meeting these demands by providing ultra-high data rate radio links \cite{shafi} -- \cite{wortecs}. Despite their very high spatial dimensions, mm-wave and THz MIMO channels are sparse. For sparse MIMO channels the number of available multi-path components (MPCs) for communications is limited and much smaller compared to the spatial channel dimensions, i.e., the MIMO channel matrix $H$ is rank deficient with its rank being much smaller than its spatial dimensions. Beamspace MIMO representation is very useful for such sparse MIMO channels because these channels are easier to learn in beamspace than in the spatial signal space \cite{sayeed} --\cite{mars}. In such cases, RF beamforming allows low cost, low power consumption, and miniaturized hardware implementation as compared to fully digital baseband processing by avoiding extra RF chains \cite{mainpap}, \cite{mt}. In the RF beamforming, the beamforming weights are implemented using digitally-controlled analog phase-shifters at the RF stage or the intermediate-frequency (IF) stage. The RF beamforming also corresponds to the first stage of the hybrid beamforming \cite{mainpap}, \cite{mt}.

The carbon footprint of information and communication technology (ICT) systems was as large as that of global air travel way back in 2008 \cite{energy} and deployment of these systems continues to outpace forecasts. Therefore, it is important to save beam training time and thereby the power consumption as much as possible. In this direction, a memory-assisted statistically ranked (MarS) algorithm was proposed in \cite{mars} to exploit the statistics of past beam training data for reducing the RF beam training overheads to a statistically minimum value. MarS not only saves the RF beam training time but also increases the robustness against the receiver noise because it was shown in \cite{ccwc} and \cite{wcsp} that a reduction in the number of beam tests increases the robustness against the receiver noise by reducing the probability of false beam selections. In \cite{mars}, the information-theoretic notions of entropy and relative entropy from \cite{shannon} were employed in the form of `beam entropy' and `relative beam entropy', respectively, for  representing angular or beam statistics of a sparse MIMO channel to choose MarS over hybrid RF beam training. The beam entropy $E$ is defined below by equation (\ref{entropy}), 
\begin{equation}
\label{entropy}
\centering
E=-\sum_{{i=1}}^{N}p_i\text{log}_2 (p_i),
\end{equation}
where $p_i$ is the probability of the $i^{th}$ beam to provide a communication path and $N$ is the number of Tx or Rx beams. In equation (\ref{entropy}), we choose the base of `2' for the logarithm because typically the RF/IF phase-shifters are controlled digitally. A beam entropy value depends on: (i) the size of the beamforming codebook, i.e., the number of beams and (ii) the channel randomness in beamspace, i.e., in  the angular domain, e.g., randomness of the angle of arrival (AoA) or/and the angle of departure (AoD). The larger the number of beams, the larger the maximum possible value of beam entropy. For a given beamforming codebook, beam entropy reaches its maximum value when all the beams are equiprobable, i.e., $p_1=p_2=...=p_i=...=p_{N-1}=p_N$. Also, the larger the randomness of the AoA or/and the AoD, the larger the respective beam entropy. The lower the beam randomness, the lower is the beam entropy $E$. If any of the beams has the probability of one, then the beam entropy $E$ will be zero. For a given beamforming codebook, relative beam entropy is defined as the ratio of the actual beam entropy to the maximum possible entropy for equiprobable beams. Having a range from 0 to 1, the relative beam entropy value depends only on the angular randomness of the channel and can be used as its single parameter indication in beamspace. Also, the relative beam entropy value remains unchanged even if the base for the logarithm is changed in equation (\ref{entropy}). Beam entropy values can also be helpful for the development of novel cryptographic algorithms by leveraging the channel stochasticity in beamspace for low overhead secured wireless communications. 

UMi, UMa, and RMa specifications are listed in Tables 7.2-1 and 7.2-3 of \cite{3GPP} and reproduced over here as Tables \ref{UMi}, \ref{UMa}, and \ref{RMa}, respectively, for convenience. The carrier frequency for RMa, i.e., rural deployment scenario is limited to 7 GHz and the key characteristics of this scenario is continuous wide area coverage supporting high speed vehicles.

\hspace{-10pt}
\begin{table}[ht]
	\caption{Parametres for Urban microcell (UMi) street canyon scenario} 
	\begin{tabular}{c c c} 
		\hline\hline 
		Sl. No. & Parametre & Value \\ [0.5ex] 
		\hline 
		1 & BS antenna height   & 10 m \\
		2 & UT mobility (hori. plane only) & 3km/h  \\
		3 & UT location & Outdoor and indoor \\
		4 & UT Height & 1 m\\ 
		5 & Environment & LoS and NLoS \\
		6 & Min. BS to UT distance (2D) & 10 m \\
		7 & UT distribution (horizontal) & Uniform \\
		\hline 
	\end{tabular}
	\label{UMi} 
\end{table}

\begin{table}[ht]
	\caption{Parametres for Urban macrocell (UMa) scenario} 
	\centering 
	\begin{tabular}{c c c} 
		\hline\hline 
		Sl. No. & Parametre & Value \\ [0.5ex] 
		\hline 
		1 & BS antenna height   & 25 m \\
		2 & UT mobility (hori. plane only) & 3km/h  \\
		3 & UT location & Outdoor and indoor \\
		4 & UT Height & 1 m\\ 
		5 & Environment & LoS and NLoS \\
		6 & Min. BS to UT distance (2D) & 35 m \\
		7 & UT distribution (horizontal) & Uniform \\
		\hline 
			\end{tabular}
	\label{UMa} 
\end{table}

\begin{table}[ht]
	\caption{Parametres for Rural macrocell (RMa) scenario} 
	\centering 
	\begin{tabular}{c c c} 
		\hline\hline 
		Sl. No. & Parametre & Value \\ [0.5ex] 
		\hline 
		1 & BS antenna height   & 35 m \\
		2 & UT mobility & High speed vehicles  \\
		3 & UT location & Outdoor, indoor, and in car \\
		4 & UT Height & 1.5 m\\ 
		5 & Environment & LoS and NLoS \\
		6 & Min. BS to UT distance (2D) & 35 m \\
		7 & UT distribution (horizontal) & Uniform \\
		\hline 
	\end{tabular}
	\label{RMa} 
\end{table}

In this paper, we present a novel research work to obtain typical beam entropy values for mm-wave channel models for UMi and UMa cellular scenarios for 28, 60, 73, and 100 GHz carrier frequencies and for RMa scenario for 3.5 GHz carrier frequency by using NYUSIM 5G mm-wave channel simulator \cite{nyusim}. The rest of the paper is organized as follows: Section \ref{prob} specifies the problem statement and the methodology, Section \ref{umi} presents simulation results, and Section \ref{conc} summarizes and concludes the paper. 
%

\section{Problem statement and methodology}
\label{prob}
It is useful to obtain typical beam entropy values for sparse mm-wave channels for 5G and beyond 5G cellular applications. This will provide an indication for beamspace-randomness of mm-wave cellular MIMO channels which is useful for RF beam training, novel cryptographic algorithms, and unforeseen future signal processing requirements for sparse MIMO radio communications.

Millimetre-wave channel models for cellular communications were developed for UMi, UMa, and RMa scenarios by the 3rd Generation Partnership Project (3GPP) in \cite{3GPP}. A comparison of 3GPP and NYUSIM channel models was presented in \cite{tedcomp} which showed that NYUSIM channel model has better physical basis and is more reliable for realistic simulations. In view of \cite{tedcomp}, it was decided to use NYUSIM simulator version 1.6.1 for this work.

A large number of MIMO channels were generated using NYUSIM with specifications as listed in Table \ref{specs}. As can be seen in serial number 2 of Table \ref{specs}, the bandwidth has been taken to be only 0.5 MHz because of the following reason: for maximum allowable bandwidth of 800 MHz with 0.5 kHz sub-carrier spacing, i.e., with 1600 sub-carriers, simulations were performed. The MIMO channel matrices were analysed for all 1600 sub-carriers using singular value decomposition (SVD) and all the sub-carriers had extremlely close singular values with the differences being in the third or fourth decimal digit. Since beam entropy characterises the channel randomness in beamspace, the beam entropy values will almost surely be the same or extremely close for all the sub-carriers. Therefore, to save computational load and time, it was decided to consider only one sub-carrier of 500 kHz bandwidth for further work reported in this paper.

\begin{table}[ht]
	\caption{Parametres common to NYUSIM simulations} 
	\centering 
	\begin{tabular}{c c c} 
		\hline\hline 
		Sl. No. & Parametre & Value \\ [0.5ex] 
		\hline 
		1 & Range & 10-500 m \\ 
		2 & RF Bandwidth & 500 kHz \\
		3 & Tx Power & 30 dBm \\
		4 & Base Station Height & 35m (RMa only)  \\
		5 & Barometric Pressure & 1013.25 mbar \\
		6 & Humidity & 50 $\%$ \\ 
		7 & Temperature  & 20 $\deg$ \\ 
		8 & Rain Rate & 0 mm/hr\\
		9 & Polarisation & Co-pol\\
		10 & Foliage loss & Nil\\
		11 & Tx Array Type & ULA\\
		12 & Rx Array Type & ULA\\
		13 & No. of Tx ULA elements & 256\\
		14 & No. of Rx ULA elements & 16\\
		15 & Tx ULA spacing & Half wavelength\\	
		16 & Rx ULA spacing & Half wavelength\\	
			17 & Environment & Line of Sight (LoS)\\		
		\hline 
	\end{tabular}
	\label{specs} 
\end{table}

As noted in serial numbers 11 to 14 of Table \ref{specs}, we take Tx and Rx antenna array to be of uniform linear array (ULA) type for simplicity. As a typical representation of base station (BS) and mobile user equipment (UE), we set the number of ULA elements to be 256 and 16 at the BS transmitter and the UE receiver, respectively, because a BS can accommodate a larger antenna array than a mobile UE with a small form-factor.

The remaining fields not recorded in Table \ref{specs} were left to their default values in NYUSIM v 1.6.1. UMa and UMi scenarios were simulated for the four frequencies, viz. 28, 60, 73, and 100 GHz, to generate $16 ~\text{x}~ 256$ MIMO matrices. RMa scenario was simulated for the carrier frequency of 3.5 GHz since it is one of the allocated bands for the cellular communications. The resultant MIMO matrices were used to perform exhaustive\footnote{Equivalently, multi-level RF beam training algorithm can also be used.} RF beam training with discrete Fourier transform (DFT) beamforming codebooks at BS Tx and UE Rx. DFT beamforming codebooks were chosen for the following reasons: (i) minimum inter-beam coupling, i.e., for DFT beamforming codebooks, at the main response axis (MRA) of one beam all other beams have their nulls. (ii) DFT beamforming codebooks also correspond to the left and right unitary matrices obtained after SVD decomposition of a MIMO channel matrix [Chapter 7 of \cite{tsebook}]. After the exhaustive RF beam training results were obtained for a large number of MIMO channel realisations, BS Tx and UE Rx beam entropy values were calculated as per equation (\ref{entropy}).
\section{Simulation results} 
\label{umi}

20,000 random $16 ~\text{x}~ 256$ MIMO channel realisations corresponding to 20,000 random user locations were obtained using NYUSIM v 1.6.1 for each of the four frequencies, viz, 28, 60, 70, and 100 GHz, for UMi and UMa cellular scenarios. In all eight cases, the MIMO channel matrices were consistently of rank one, having channel condition numbers of about 166 dBs, indicating one dominant cluster with much weaker non-dominant clusters in the cellular mm-wave environment. The typical Rx and Tx beam entropies were about 3.75 and 7.62, varying in third and fourth decimal places for different scenarios and different frequencies. For a $16 ~\text{x}~ 256$ MIMO channel using DFT beamforming codebooks at Rx and Tx, the maximum possible Rx and Tx beam entropies are 4 and 8, respectively. So, typical Rx and Tx relative beam entropies are 0.9375 and 0.953, respectively.

In further simulations, the Tx power was increased from 30 dBm to 60 dBm for the 28 GHz UMi scenario. Despite the increased Tx power, the same values of Rx and Tx beam entropies were obtained as with Tx power of 30 dBm. In other words, we do not observe changes in Rx and Tx beam entropy values with increase in Tx power. This reinforces the understanding that the non-dominant clusters are too weak compared to the dominant clusters so that even a thousand fold increase in the Tx power also does not lead to a creation of new dominant clusters.

With MIMO channel dimensions changed from $16 ~\text{x}~ 256$ to $8 ~\text{x}~ 128$ for the 28 GHz UMi scenario, we obtained Rx and Tx beam entropy values of 2.7927 and 6.6759 for 8315 random channel realisations. These beam entropy values correspond to Rx and Tx relative beam entropy values of 0.9309 and 0.9537, respectively. These relative beam entropy values are also very close to those obtained for simulations with $16 ~\text{x}~ 256$ MIMO channels.

\begin{figure}[!t]
	\hspace{-10pt}
	\includegraphics[width=3.5in]{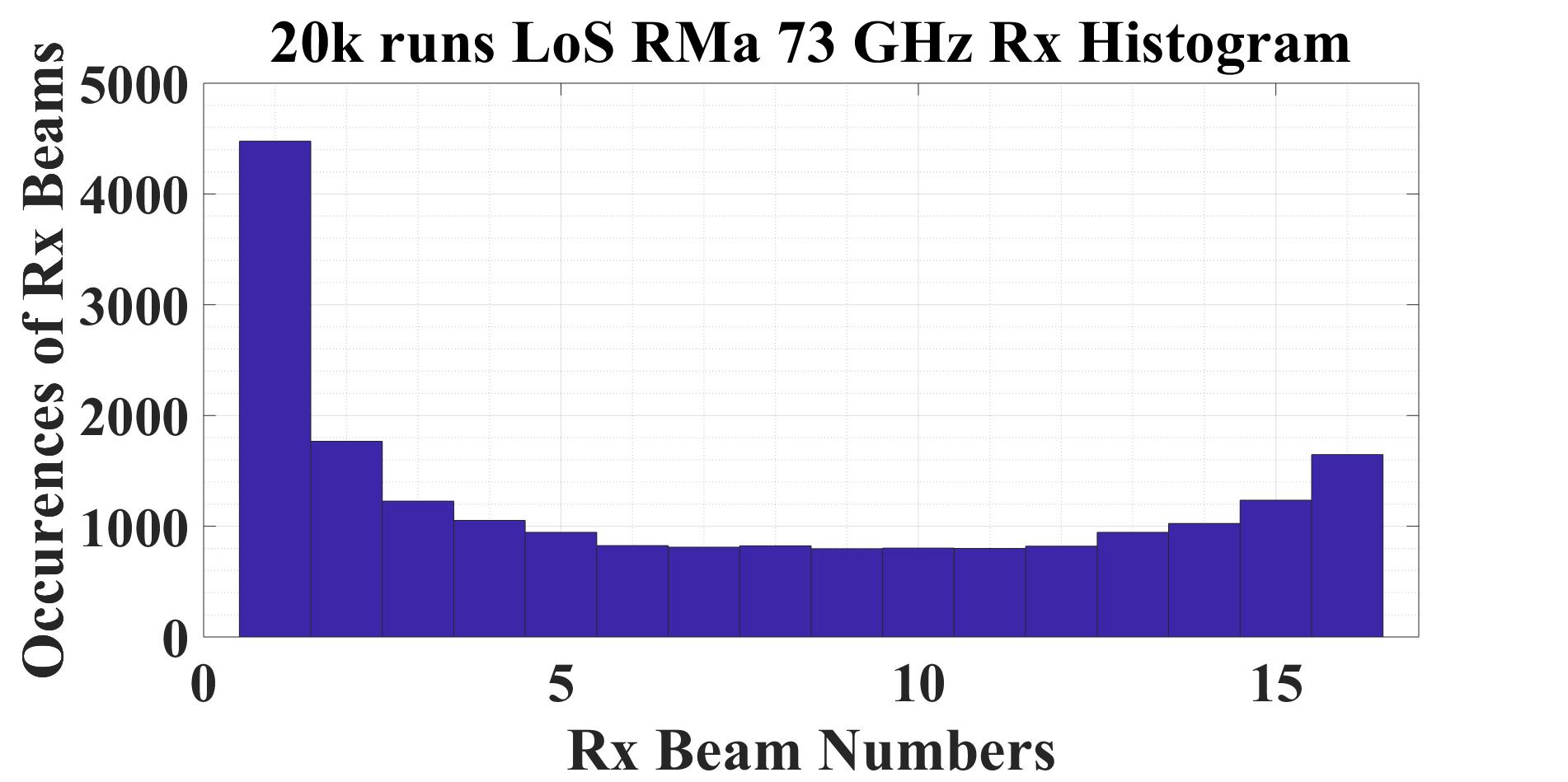}
	\caption{Rx beam histogram for 28 GHz UMi scenario}
	\label{hist}
\end{figure} 

In Figure \ref{hist}, we can see the Rx beam histogram for the 28 GHz UMi scenario with MIMO channel dimension of $8 ~\text{x}~ 128$ for 8315 random channel realisations. As stated above, the Rx beam entropy for this case was 2.7927 corresponding to a relative beam entropy value of 0.93. We can see that the central beam corresponding to the Rx ULA broadside have almost equal numbers of occurrences and therefore probabilites because if we normalise this histogram by the total number of runs we get the probability mass function (PMF) of Rx beams. We also note from Table \ref{UMi} that the UE or UT distribution is uniform which is a good match. We observe that relative beam entropy values remain almost the same even with change in MIMO dimensions. 


We note that the maximum carrier frequency for RMa scenario is 7 GHz for continuous wide area coverage supporting high speed vehicles. We choose 3.5 GHz since it is one of the allocated bands. RMa scenario was simulated for the carrier frequency of 3.5 GHz with $4 ~\text{x}~ 128$ MIMO channel. For 20,000 random channel realisations and 30 dBm Tx power, the Rx and Tx beam entropy were 1.851 and 6.693, corresponding to Rx and Tx relative beam entropy values of 0.926 and 0.956, respectively. For 60 dBm Tx power, the Rx and Tx relative beam entropy values were 0.926 and 0.956, respectively, similar to UMi and UMa scenarios. 

The cellular mm-wave channels have high relative beam entropy values because of the uniform UT distributions as shown in Tables \ref{UMi}, \ref{UMa}, and \ref{RMa}. From the view point of application of MarS and hybrid RF beam training algorithms \cite{mars}, the high relative beam entropy values are of little advantage since the different beams are almost equally likely to provide a communication path. However, such high relative beam entropy values may be useful for developing novel cryptographic key generation algorithms which  leverage the high randomness of beams, probably aided by a beam-tracking algorithm \cite{bt}. For the cases of very narrow beams, the possibility of eavesdropping is not very high due to the limited spatial spread of the electromagnetic energy. The beam entropy values can potentially serve as a useful metric for novel access or radio resource allocation protocols. Beam entropy metric can also be useful for unified communications and sensing based on the common radio hardware, e.g., automotive and industrial applications. As an example, a beam entropy value, based on the memory-assisted beam statistics, can be useful for sensing applications or sensor systems where a search needs to be performed in a given angular space, say searching a target on ground or searching an aircraft in the sky, etc. Currently, sequential scans are used, e.g., for military applications where the beam statistics data is not available and target distribution is expected to be typically uniform in the search zone. With emerging sensing applications in consumer and civilian sectors, e.g., automotive radars, etc., a tailored form of MarS could be useful to save search time and resources for non-uniform target distributions. 




%

\section{Conclusions}
\label{conc}
NYUSIM simulations resulted in typical Rx and Tx relative beam entropy values of 0.938 and 0.953, respectively, for 5G cellular mm-wave channels. Such high relative beam entropy values indicate a little advantage in employing MarS RF beam training algorithms for cellular mm-wave channels. There is a potential case for the possibility of low overhead secured communications. 

Relative beam entropy values remained almost unchanged with increase in Tx power and MIMO dimensions. Relative beam entropy values increased very gradually with an increase in the number of channel realisations used for calculating beam entropy values.

This work used one and the same value for single element antenna gain for all frequencies, while for a given form-factor higher frequency antennas should achieve higher gains. It is an interesting future work to obtain typical beam entropy values for different frequencies with frequency variant single element antenna gains. It will also be useful to investigate typical beam entropy values for higher frequencies, especially towards 300 GHz, and indoor applications, e.g., virtual reality use case of EU Horizon 2020 WORTECS project \cite{wvr}, \cite{wr}.

\section*{Acknowledgment}
This work has received funding from the European Union's Horizon 2020 research and innovation programme under grant agreement No. 761329 (WORTECS).

The first author would like to thank Shihao Ju of NYU WIRELESS and NYU Tandon School of Engineering, New York University, USA for NYUSIM support and Klaus Tittelbach-Helmrich for reviewing the manuscript.




\begin{thebibliography}{1}

\bibitem{cisco} 
Cisco white paper, ``Cisco Visual Networking Index: Forecast and Trends, 2017–2022 ," Cisco Document ID: 1551296909190103, Feb. 2019, [Online]. Available: https://www.cisco.com/c/en/us/solutions/collateral/service-provider/visual-networking-index-vni/white-paper-c11-741490.html [Accessed: 20-Jun-2019].  

\bibitem{shafi}  
M.~Shafi et al., ``5G: A tutorial overview of standards, trial, challenges, deployment, and practice," \emph{IEEE J. Sel. Areas Commun.}, vol. 35, no. 6, pp. 1201-1221, Jun. 2017.

\bibitem{rangan}  
S.~Rangan, T.~S.~Rappaport, and E.~Erkip, ``Millimeter-wave cellular wireless networks: Potentials and challenges," \emph{Proc. IEEE.}, vol. 12, no. 3, pp. 366-385, Mar. 2014.

\bibitem{2014}  
I.~F.~Akyildiz, J.~M.~Jornet, and C.~Han, ``Terahertz Band: Next
Frontier for Wireless Communications," \emph{Phys. Commun.}, vol. 12, no. 2, pp. 16-32, Sept. 2014.

\bibitem{3GPP}  
3GPP, ``Study on channel model for frequencies from 0.5 to 100 GHz (Release 15)," 3rd Generation Partnership
Project (3GPP), TR 38.901 V15.0.0, June 2018. [Online]. Available:
https://portal.3gpp.org/desktopmodules/Specifications/Specificati\\onDetails.aspx?specificationId=3173 [Accessed: 16 June 2019].	

\bibitem{wortecs} EU WORTECS Consortium, ``European Union's Horizon 2020 research and innovation programme under grant agreement No 761329 WORTECS project website," [Online]. Available: https://wortecs.eurestools.eu/ [Accessed: 20-Jun-2019].
	
 \bibitem{sayeed}  
	A.~M.~Sayeed, ``Deconstructing multiantenna fading channels," \emph{IEEE Transactions on Signal Processing}, vol. 50, no. 10, pp. 2563-2579, Oct. 2002. 
	
\bibitem{bradjournal}  
	J.~Brady, N.~Behdad, and A.~Sayeed, ``Beamspace MIMO for Millimeter-Wave Communications: System Architecture, Modeling, Analysis, and Measurements," \emph{IEEE Transactions on Antennas and Propagation}, vol. 61, no. 7,  pp. 3814-3827, Jul. 2013.
	
\bibitem{bradsay}  
	A.~Sayeed and J.~Brady, ``Beamspace MIMO for high-dimensional multiuser communication at millimeter-wave frequencies," \emph{IEEE Global Communications Conference (GLOBECOM)}, Atlanta, GA, Dec. 2013, pp. 3679-3684. 
	
\bibitem{bradsong}  
	G.~H.~Song, J.~Brady, and A.~Sayeed, ``Beamspace MIMO transceivers for low-complexity and near-optimal communication at mm-wave frequencies," \emph{IEEE International Conference on Acoustics, Speech and Signal Processing (ICASSP)}, Vancouver, BC, May 2013, pp. 4394-4398.
	
	\bibitem{mars}  
	K.~K.~Tiwari, E.~Grass, J. S. Thompson, and R.~Kraemer, ``Memory-assisted Statistically-ranked RF Beam Training Algorithm for Sparse MIMO," \emph{IEEE Global Communications Conference (GLOBECOM)}, under review, arXiv preprint https://arxiv.org/abs/1906.01719.
	
	\bibitem{mainpap}  
	A. Alkhateeb, G. Leus, and R. W. Heath, ``Limited feedback hybrid precoding for multi-user millimeter wave systems," \emph{IEEE Transactions on Wireless Communications}, vol. 14, no. 11, pp. 6481-6494, Nov. 2015.
	
\bibitem{mt}
	K. K. Tiwari, ``Beamforming Techniques for Millimetre Wave Communications," \emph{ M. Sc. thesis, matriculation number s1113092}, The University of Edinburgh, U.K., Aug. 2017.
	
	\bibitem{energy}  
	G.~P.~Fettweis and E.~Zimmermann, ``ICT Energy Consumption-Trends and Challenges," \emph{The 11th International Symposium on Wireless Personal Multimedia Communications (WPMC)}, Lapland, Finland, Sep. 2008, pp. 1-4. 
	
		\bibitem{ccwc}  
	K.~K.~Tiwari, E.~Grass, and R.~Kraemer, ``Noise performance of Orthogonal RF beamforming for THz Radio Communications," \emph{The 9th IEEE Annual Computing and Communication Workshop and Conference (CCWC)}, Las Vegas, USA, Jan. 2019, pp. 746-751.
	
	\bibitem{wcsp}  
	K.~K.~Tiwari, J.~S.~Thompson, and E.~Grass, ``Noise performance of Orthogonal RF beamforming for millimetre wave massive MIMO communication systems," \emph{The 10th International Conference of Wireless Communications and Signal Processing (WCSP)}, Hangzhou, China, Oct. 2018, pp. 1-7.
	
	  \bibitem{shannon}  
	C. E. Shannon and W. Weaver, \emph{The Mathematical Theory of Communication}, The University of Illinois Press, 1971.
	
\bibitem{nyusim}  
	S. Sun, G. R. MacCartney and T. S. Rappaport, ``A novel millimeter-wave channel simulator and applications for 5G wireless communications," \emph{2017 IEEE International Conference on Communications (ICC)}, Paris, May 2017, pp. 1-7.
	
	
\bibitem{tedcomp}  
T. S. Rappaport, S. Sun and M. Shafi, ``Investigation and Comparison of 3GPP and NYUSIM Channel Models for 5G Wireless Communications," \emph{2017 IEEE 86th Vehicular Technology Conference (VTC-Fall)}, Toronto, ON, 2017, pp. 1-5.

  \bibitem{tsebook}  
D.~Tse and P.~Viswanath, \emph{Fundamentals of Wireless Communication}, Cambridge University Press, 2005.

\bibitem{bt}  
K.~K.~Tiwari, V.~Sark, E.~Grass, and R.~Kraemer, ``Monopulse-based THz Beam Tracking for Indoor Virtual Reality Applications," \emph{ 24th ITG Fachtagung Mobilkommunikation Technologien und Anwendungen}, Osnabr\"{u}ck, Germany, May 2019, pp. 10-13.

\bibitem{wvr} M.~ Badawi et. al, ``EU H2020 WORTECS Deliverable D2.3: Focus on Virtual Reality," [Online]. Available: https://wortecs.eurestools.eu/deliverables-dissemination/ [Accessed: 16 June 2019].

\bibitem{wr} G.~ Vercasson et. al, ``EU H2020 WORTECS Deliverable D3.2: Common Analog and Digital Baseband Design for Flexible Radio and Optical Transceiver," [Online]. Available: https://wortecs.eurestools.eu/deliverables-dissemination/ [Accessed: 16 June 2019].





 
 \end{thebibliography}
%


\end{document}